\documentclass[twocolumn,prl]{revtex4}
\usepackage{graphicx}

\newcommand{\bra}[1]{\left\langle{#1}\right|}
\newcommand{\ket}[1]{\left|{#1}\right\rangle}

\begin{document}

\title{$d$-wave resonating valence bond states of fermionic atoms in optical lattices}

\author{Simon Trebst$^{1,2,3}$,  Ulrich Schollw\"ock$^{4}$, Matthias Troyer$^{1}$, Peter Zoller$^{5}$}

\affiliation{$^{(1)}$Theoretische Physik, Eidgen\"ossische
Technische Hochschule
 Z\"urich, CH-8093 Z\"urich, Switzerland}
\affiliation{$^{(2)}$Computational Laboratory,  Eidgen\"ossische
Technische Hochschule
 Z\"urich, CH-8092 Z\"urich, Switzerland}
\affiliation{$^{(3)}$ Microsoft Research and Kavli Institute for Theoretical Physics, University of California, Santa Barbara, CA 93106, USA}
\affiliation{$^{(4)}$ Institut f\"ur Theoretische Physik C, RWTH
Aachen, D-52056 Aachen, Germany} 
\affiliation{$^{(5)}$Institute
for Theoretical Physics, University of  Innsbruck, and \\Institute
for Quantum Optics and Quantum Information of the Austrian Academy
of Science, 6020 Innsbruck, Austria}
\date{\today}

\begin{abstract}
We study controlled generation and measurement of superfluid
$d$-wave resonating valence bond (RVB) states of fermionic atoms
in 2D optical lattices. Starting from loading spatial and spin
patterns of atoms in optical superlattices as pure quantum states
from a Fermi gas, we adiabatically transform this state to an RVB
state by change of the lattice parameters. Results of exact
time-dependent numerical studies for ladders systems are
presented, suggesting generation of RVB states on timescale
smaller than typical experimental decoherence times.
\end{abstract}


\maketitle


Resonating valence bond (RVB) states, in
which electrons are paired into short-range singlets by a purely
electronic mechanism, were originally introduced
\cite{Anderson:73} as wave functions for insulating spin liquid
ground states of Mott insulators. Shortly after the observation of
high-temperature superconductivity in the cuprates
\cite{Bednorz:86}, Anderson conjectured that they might be
described by doped RVB states \cite{Anderson:87}, in which the RVB
pairs of the insulating state become mobile and superconducting
upon doping.
Since the predicted exotic $d$-wave pairing symmetry
\cite{Kotliar:88} of the RVB pairs was confirmed experimentally in
the cuprate superconductors \cite{Tsuei:00}, the RVB scenario
remains as one of the most promising contenders for the theory of
high temperature superconductivity \cite{RVB:04}. While the ground
state of the lightly doped two-dimensional Hubbard model is still
unknown, the $d$-wave RVB scenario has been confirmed for $t$-$J$
and Hubbard models of coupled plaquettes \cite{fye,Altman} and ladders
(consisting of two coupled chains) \cite{Tsunetsugu:94} by
numerical simulations. The key question is whether the RVB state of weakly coupled ladders  \cite{Tsunetsugu:95} or plaquettes  \cite{Altman} is connected to a
possible RVB ground state of the square lattice Hubbard model: does the RVB state survive when the inter-plaquette (ladder) coupling is increased and becomes the same as the intra-plaquette (ladder) coupling, at which point we have a uniform square lattice?

We propose to address this question using cold atoms loaded in optical lattices, which allow the realization of
Hubbard models with controllable parameters, promising an entirely
new avenue in the study of strongly correlated systems
\cite{Jaksch:98,Hofstetter:02,Lew,Sorensen,Buechler:05,Bloch,fermilattice,Kohl:04}.
In particular, formation of RVB ground states with fermionic atoms
in a 2D geometry is one of the ultimate challenges
\cite{Hofstetter:02}. However, d-wave pair binding energies are
two orders of magnitude smaller than the hopping amplitudes
\cite{Altman} for atoms in an optical lattice which are typically
in the kHz range: this poses a significant experimental challenge
in terms of temperature requirements {\em etc.}, which will be
difficult to meet. Instead we propose here the formation of RVB
states by loading spatial and spin patterns of atoms in
optical superlattices as {\em pure quantum states} from a reservoir of a
quantum degenerate Fermi gas \cite{fermigas}, and {\em
adiabatically transforming} this state to an RVB state by change
of the lattice parameters. An initial lattice configuration is
designed in such a way that the desired initial state corresponds
to the ground state of the deformed lattice in the form of a
simple many-body product state, with a large excitation gap. This
makes the preparation of these states robust against
imperfections. In choosing a protocol for the deformation of the
lattice to achieve the RVB state, the challenge is to select a
parameter path, which minimizes the number of possible avoided 
crossings, and which sets the time scale for the adiabatic transformation,
reminiscent of discussions in adiabatic quantum computing
\cite{adiabaticqc}.
While this technique is illustrated for the case of RVB states here, it might 
provide a generic procedure to generate non-trivial ground states of dilute 
fermionic gases in lattices.

In this Letter we present a detailed investigation of this scenario.
We will start with the loading of spin and spatial patterns
on decoupled plaquettes formed by optical superlattices representing
different atomic dopings, and discuss requirements for the formation of
RVB plaquette states, and possible experimental signatures
demonstrating $d$-wave pairing.
We present results of exact time-dependent numerical studies for ladders
systems, suggesting that RVB states can be generated on a timescale
smaller than the typical decoherence time of atoms in optical lattices.
Since the numerical simulation of 2D strongly correlated systems belongs
to the class of non-deterministic polynomial-time hard (NP-hard) 
computational problems \cite{TroyerWiese}, we are not able to provide 
answers for the coupling of plaquettes or ladders to probe the ground
state of the 2D Hubbard model, but pose this as an important problem 
left for experiments to be solved.

\begin{figure}[t]
  \includegraphics[width=86mm]{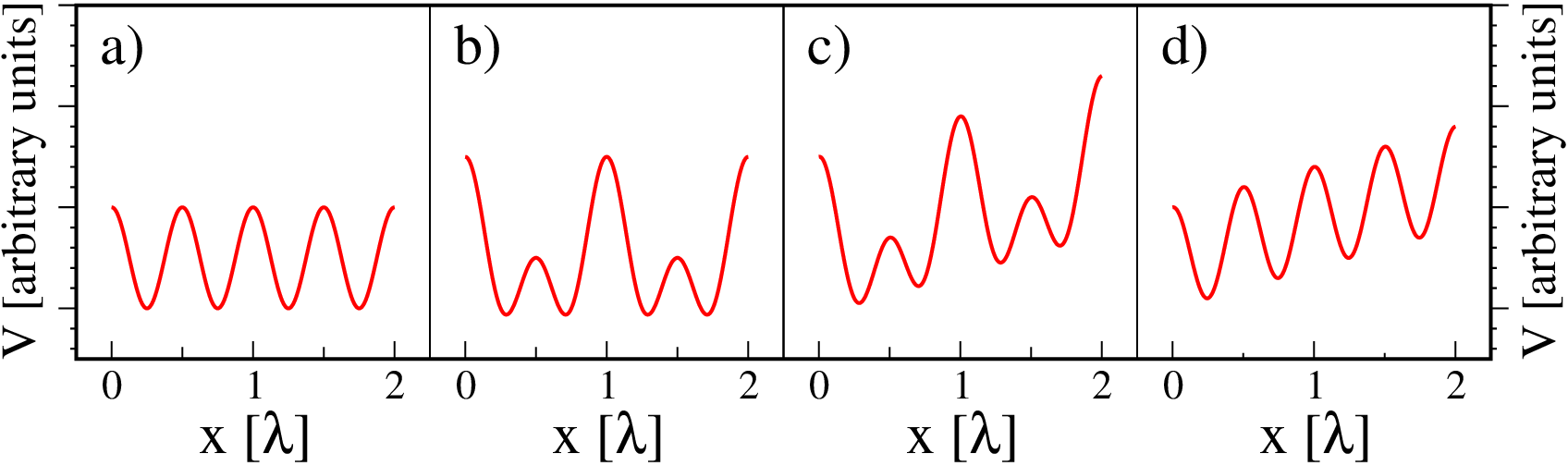}
  \caption{
    Optical potentials generated by two counterpropagating laser beams
    with wave vector $k=2\pi/\lambda$.
    a) $V=V_0 \sin(kx)^2$;
    b) $V=V_0 \sin(kx)^2 + V_1 \sin(kx/2)^2$;
    c) $V=V_0 \sin(kx)^2 + V_1 \sin(kx/2)^2 + V_2x$;
    d) $V=V_0 \sin(kx)^2 + V_2x$.}
  \label{Fig:optical_potential}
\end{figure}

{\em Optical lattices --} We start by summarizing the basic tools
available for construction of 1D ladders and 2D square lattices
with optical potentials
\cite{Jaksch:98,Hofstetter:02,Lew,Sorensen,Buechler:05}.
Counterpropagating laser beams along the $x_i$ ($i=1,2,3$)
directions realize a periodic optical potential $V_0(\vec x) =
\sum_{i} V_{0i} \sin (k x_i)^2$ with wave vector $k=2
\pi/\lambda$, optical wave length $\lambda$, and lattice depths
$V_{0i}$ controlled by the laser intensities and atomic alternating 
current (AC) polarizabilities. 
These potentials are additive, provided the
lasers generating the lattice have slightly different optical
frequencies, which are easily generated as sidebands from the
original laser beam, so that the interference terms average out.
Furthermore, interference of two lasers with
 angles $\pm\theta$ between their propagation
directions give rise to potentials of the form $V_{1i}  \sin (k'
x_i + \Phi)^2$ with effective wave vector $k'=k \cos \theta$ and
displaced by a phase $\Phi$. This allows us to construct in
particular potentials $V_1(\vec x) = \sum_{i} V_{1i} \sin (k x_i
/2)^2$. Adding these various potentials gives rise to superlattice
structures. We can also add potentials in the form of linear
ramps, $V_2(\vec x) = \sum_{i} V_{2i} x_i$ by sitting in the wing
of a focused laser beam, as in far-off resonance laser traps.
Fig.~\ref{Fig:optical_potential} illustrates lattice
configurations which can be generated in this way by varying the
intensity parameters $V_{0,1,2i}$ which will be employed below for
the construction of $d$-wave pairing.

The dynamics of cold atoms confined to these optical potentials is
described by a Hubbard model
\cite{Jaksch:98,Hofstetter:02,Lew,Sorensen,Buechler:05}
 \[\label{Hubbard} H = - \sum_{\langle
i,j \rangle, \sigma} t_{ij} \left( c^{\dagger}_{i\sigma}
c^{\phantom\dagger}_{j\sigma} + h.c. \right) +\, U \sum_i
n_{i\uparrow}n_{i\downarrow} + \sum_{i,\sigma} \mu_i\, n_{i\sigma}
\,,
\]
where the $c_{i,\sigma}$ are fermionic annihilation operators with
spin $\sigma$ and $n_{i,\sigma} = c^{\dagger}_{i,\sigma}
c^{\phantom\dagger}_{i,\sigma}$ is the particle number on site
$i$. The $t_{ij}$ are spatially dependent hopping matrix elements
connecting neighboring sites $i$ and $j$, and the $\mu_i$ are site
offsets, as determined by the superlattice structure (see
Fig.~\ref{Fig:optical_potential}). The collisional repulsion $U$
between the atoms can be controlled by Feshbach resonances
\cite{fermigas,Feshbach}. In writing the Hubbard model we have
assumed that the optical potentials are spin independent, which is
the usual case of Alkali atoms in their ground state
\cite{Jaksch:98}.


{\em Plaquette RVB States --} The motion of the atoms can be
confined to a 2D lattice by a strong transverse optical potential
$V_{0z}$. Employing superlattices we can generate double well
potentials corresponding to decoupled plaquettes. Our first goal
will be to study atomic dynamics on these uncoupled plaquettes,
and in particular the generation of RVB (ground) states. The
strategy is to (i) deform the optical lattice on each plaquette so
that the corresponding ground state has a {\em simple product
form}, $|\psi\rangle=\prod_{i\sigma} c^{\dagger}_{i\sigma}
|0\rangle$, and (ii) adiabatically transform the lattice into an
unperturbed plaquette, so that the final ground state is the
desired RVB state. To prepare such a pure initial state we load
atoms from a reservoir of a quantum degenerate Fermi gas via a
coherent or dissipative Raman process  into the optical lattice
\cite{Rabl:03}. By choosing an appropriate pattern of site offsets
in the lattice, we ensure only atoms are transferred which match
the energy conservation condition. As shown in \cite{Rabl:03},
this allows us to {\em filter out} from an initial finite entropy
ensemble a pure spin and spatial pattern of atoms.

\begin{figure}[t]
  \includegraphics[height=24.5mm]{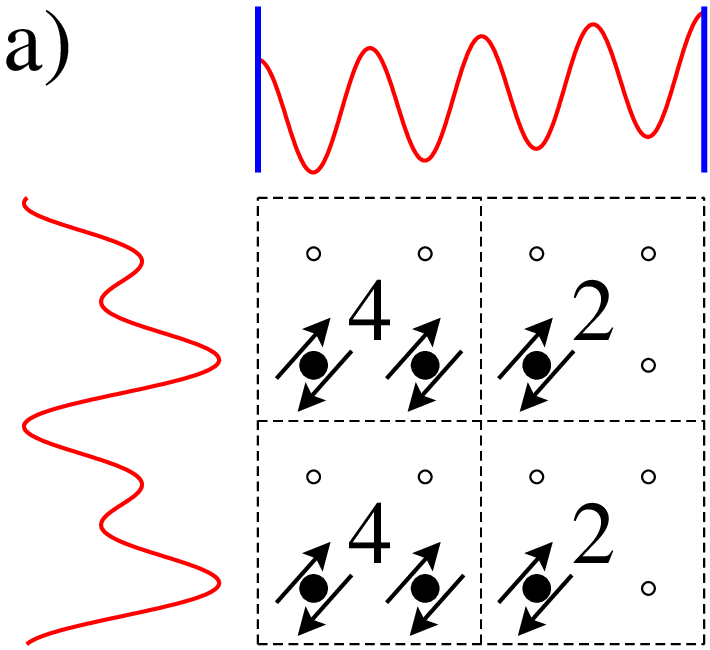}
  \includegraphics[width=30mm]{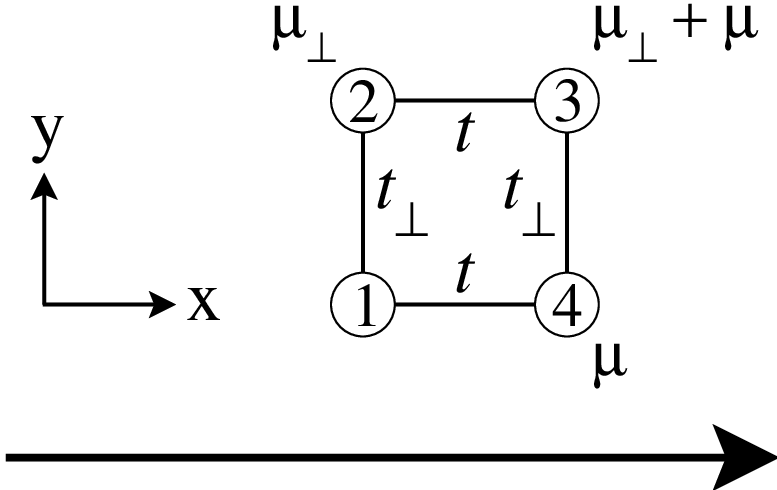}
  \includegraphics[height=24.5mm]{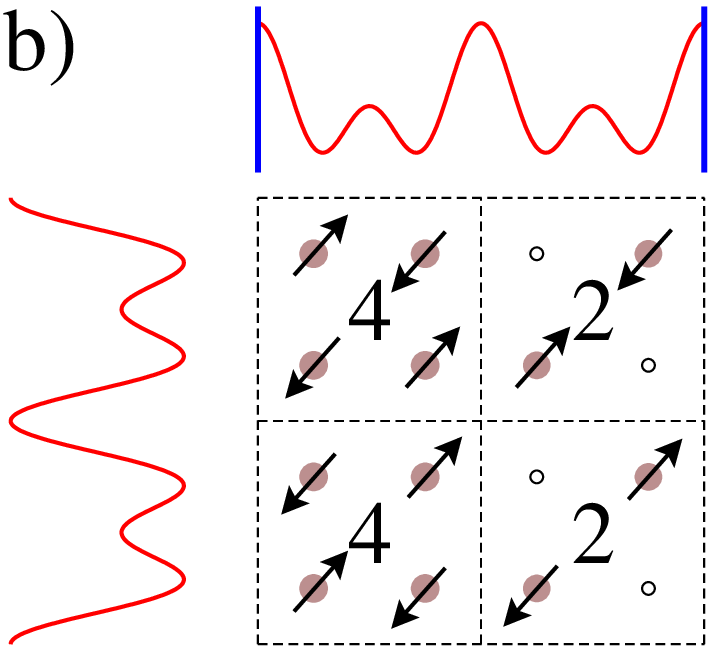}
  \caption{
    Schematic illustration of the adiabatic protocol that generates RVB
    states on two decoupled plaquettes with 4 atoms (left) and 2 atoms (right).
    The protocol sequence is: 1) $t\rightarrow1$; 2) $\mu\rightarrow0$;
    3) $\mu_{\perp}\rightarrow 0$.
    Optical potentials are sketched for $x$ and $y$-directions and 
    the large (intermediate, small) circles indicate two (one, zero) atoms on a site.
    }
  \label{Fig:StartConfiguration}
\end{figure}

We consider preparation of one plaquette with 4 atoms
(half-filling) and one plaquette with 2 atoms. The initial optical
lattice for these two cases is described in
Fig.~\ref{Fig:StartConfiguration}a), which amounts in the limit
$t_\perp \ll \mu_\perp$ to a preparation of the states with 4 (2)
atoms in a product state of the form
\begin{equation}
    \ket{4}^{(0)}  =  c^{\dagger}_{1\uparrow}c^{\dagger}_{1\downarrow}
                                           c^{\dagger}_{4\uparrow}c^{\dagger}_{4\downarrow} \ket{0} \;, \quad \quad
    \ket{2}^{(0)} = c^{\dagger}_{1\uparrow}c^{\dagger}_{1\downarrow} \ket{0} \;,
\end{equation}
where the indices 1 and 3 denote the wells along the $x$-direction
as illustrated in Fig.~\ref{Fig:StartConfiguration}.


An adiabatic protocol that allows to transform this initial product
state to an RVB plaquette state is presented in
Fig.~\ref{Fig:StartConfiguration} using the transformations of
optical lattices illustrated in Fig.~\ref{Fig:optical_potential}.
Full plaquette symmetry is restored by coupling two wells first along
the $x$-direction with their fully depleted counterparts along the
$y$-direction.
The initial optical lattice breaks reflection symmetry along the
$x$-direction which upon coupling by increasing $t$
($V_{0x} \rightarrow 0$) enforces that the
state $\ket{2}^{(0)}$ adiabatically connects only to the dimer state 
which is antisymmetric under spin exchange and symmetric under reflection.
Eliminating the chemical potential shifts $\mu$ ($V_{2x}\rightarrow0$)
and $\mu_{\perp}$ ($\Phi\rightarrow0$) one can adiabatically prepare
the ground states of a 4-site plaquette state.
For 4 atoms it is given by
\begin{equation}
    \ket{4}  \approx  \frac{1}{\sqrt{2}} \left( s^{\dagger}_{1,2} s^{\dagger}_{3,4}
                    - s^{\dagger}_{1,4} s^{\dagger}_{2,3} \right) \ket{0} \;,
\end{equation}
where we have only written the dominant terms omitting states with
local double occupancy, and 
 $s^{\dagger}_{i,j} = \left( c^{\dagger}_{i\uparrow}c^{\dagger}_{j\downarrow} -  c^{\dagger}_{i\downarrow}c^{\dagger}_{j\uparrow}\right)/\sqrt{2}$
is a singlet state formed on sites $i$ and $j$. 
On the other plaquette the 2 atoms form a state 
\begin{equation}
     \ket{2}   \approx \left[ \frac{1}{\sqrt{8}} \left( s^{\dagger}_{1,2} + s^{\dagger}_{3,4}
                    + s^{\dagger}_{1,4} + s^{\dagger}_{2,3} \right)  
                    + \frac{1}{2} \left( s^{\dagger}_{1,3} + s^{\dagger}_{2,4} \right) \right] \ket{0} \;.
\end{equation}
While the state $\ket{2}$ has $s$-wave symmetry with respect to the 
vacuum, it has $d_{x^2-y^2}$-symmetry with respect to the ground 
state $\ket{4}$ at half-filling, as the bosonic $d$-wave hole pair 
operator $\Delta = \left( s_{1,2} + s_{3,4} - s_{1,4} - s_{2,3} \right)/2$ 
has a non-vanishing contribution $\bra{2} \Delta \ket{4} \neq 0$. 
The state $\ket{2}$ is therefore referred to as the $d$-wave RVB state 
(of the hole pair).
The whole transformation is protected by a finite gap
that sets the time scale for an adiabatic coupling with a fidelity
larger than 0.99 to $T \approx 50/t$.


\begin{figure}[t]
  \includegraphics[width=86mm]{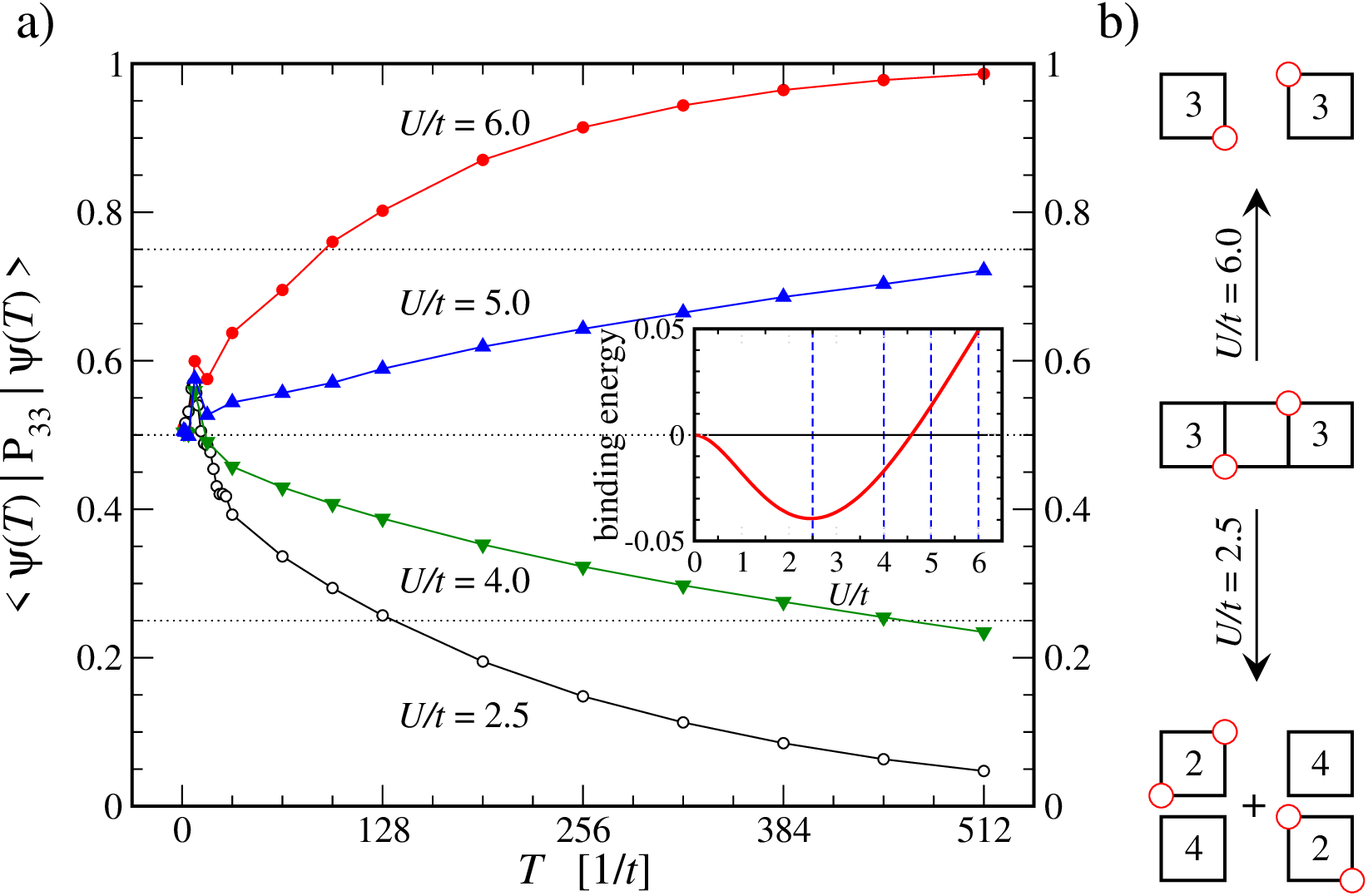}
  \caption{
    Signature for $d$-wave RVB pairing when decoupling
    two plaquettes with 6 atoms for varying onsite repulsion $U/t$.
    b) For $U/t \lesssim 4.5$ pairs are formed with a small binding
    energy shown in the inset. In the final state after sufficiently slow
    decoupling the hole pair is located on one plaquette, while for
    large repulsion $U/t \gtrsim 4.5$ the unpaired holes separate
    into 3 atoms on each plaquette (right panel).
    a) Projection of the final state onto the subspace
    with 3 atoms on each plaquette.
  }
  \label{Fig:CoupledPlaquettes_Projection}
  \vspace{-1mm}
\end{figure}

{\em Coupling of plaquettes --} Coupling and decoupling these two
plaquettes can provide a clear experimental signature for the
pairing of atoms (or holes) in these RVB states, see
Fig.~\ref{Fig:CoupledPlaquettes_Projection}. When coupling the
plaquettes the holes move to the center of the 8 coupled sites as
illustrated in Fig.~\ref{Fig:CoupledPlaquettes_Projection}b).
Subsequently decoupling the two plaquettes, the various atomic
occupation numbers on the two plaquettes are in direct
competition. If a bound state of two atoms (holes) is disfavored
by large onsite repulsion (inset of
Fig.~\ref{Fig:CoupledPlaquettes_Projection}a), the atoms will
maximize their respective kinetic energies as the plaquettes are
decoupled and the holes separate. After adiabatic decoupling there
are two plaquettes with 3 atoms each as shown in
Fig.~\ref{Fig:CoupledPlaquettes_Projection}. If a bound state is
favorable the pair hops to either one of the plaquettes, leaving
the system in a final state with an even number of atoms on both
plaquettes. Pair binding is found for $U/t \lesssim 4.5$ with a
maximum binding energy of $E_b/t \approx 0.04$ for $U/t \approx
2.5$ (see inset of Fig.~\ref{Fig:CoupledPlaquettes_Projection})
\cite{Altman}. The time scale for the adiabatic decoupling is
dominated by the small binding energy scale and found to be $T
\approx 500/t$ or $T\approx 1/2$ seconds for $^{40}$K atoms which
is well within the decoherence time of these systems \cite{Kohl:04}. 
Reversing the adiabatic protocol illustrated in
Fig.~\ref{Fig:StartConfiguration} the plaquette states after
decoupling are transformed into single-site states. For pair
binding the system ends up with 3 doubly occupied sites, while for
unbound pairs there will be two doubly occupied sites and two
sites with one atom each. Experimentally the two scenarios can be
distinguished by associating the atoms into molecules
\cite{Buechler:05,moleclattice} with the number of formed
molecules having a ratio of 3/2 respectively.

\begin{figure}[b]
  \includegraphics[width=86mm]{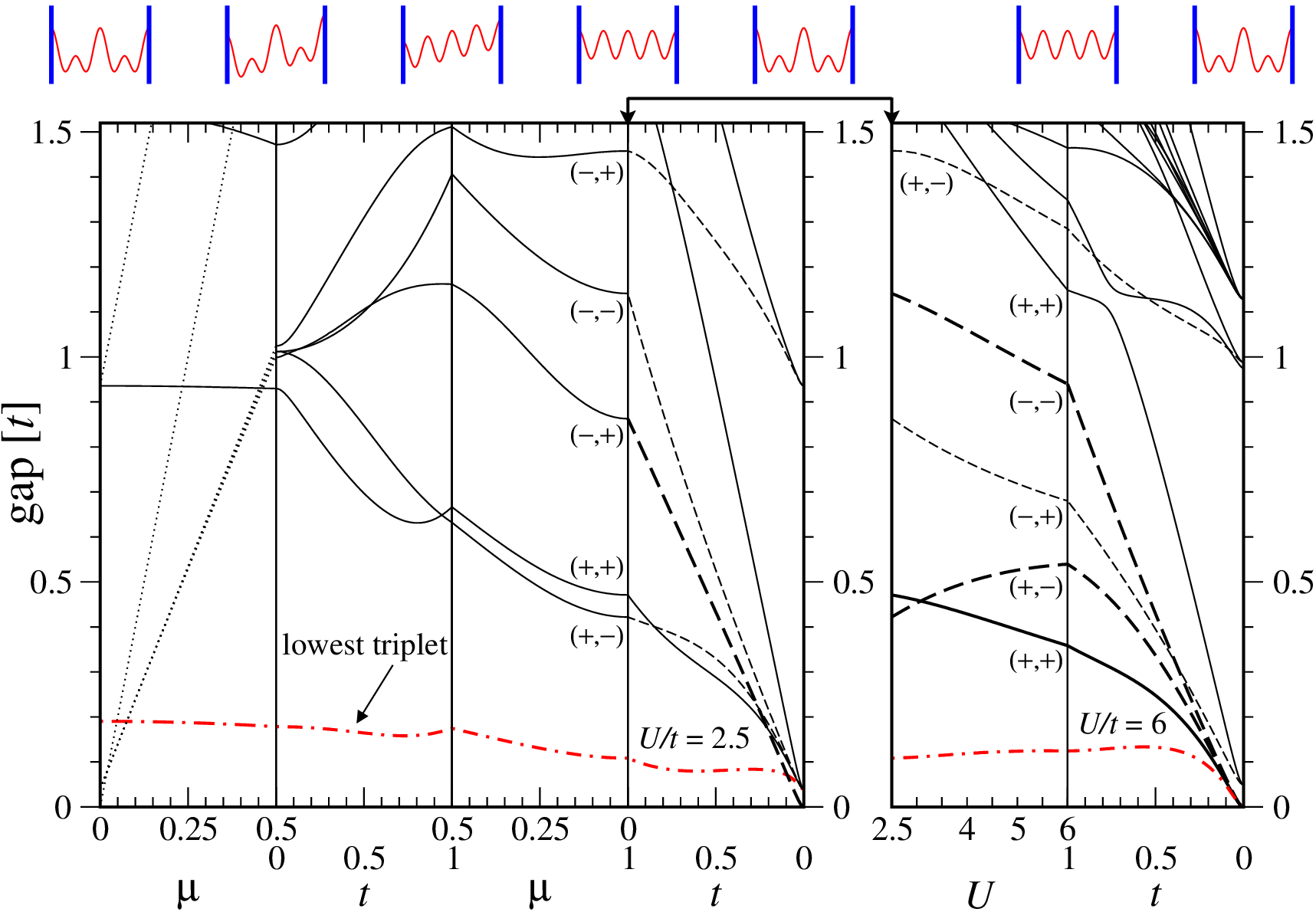}
  \caption{
        Singlet gaps for the path that couples and decouples two plaquettes
        illustrated by the optical potentials on top.
        Two plaquettes with 4 and 2 atoms are coupled for $U/t=2.5$ with a
        chemical potential gradient $\mu$ being applied along the
        $x$-direction (first three panels).
        The decoupling of plaquettes is shown for bound (fourth panel) and
        unbound pairs (right panels).
        Dashed (dotted) lines indicate states to which a transition from the
        ground state is forbidden due to symmetry constraints (particle conservation
        on the individual plaquettes).
        The  $x$- and $y$-parities are given by "+" (even) and "--" (odd).
  }
  \label{Fig:PlaquetteGaps}
\end{figure}

The detailed adiabatic protocol that couples two plaquettes is
illustrated on top of Fig.~\ref{Fig:PlaquetteGaps}. While the
final ground state of two plaquettes with 6 atoms is symmetric
with respect to the exchange of the two plaquettes, coupling two
plaquettes prepared in the states $\ket{4}$ and $\ket{2}$ will
adiabatically connect the initial state $\ket{4}\ket{2}$ to both
(anti)symmetric combinations $(\ket{4}\ket{2} \pm
\ket{2}\ket{4})/\sqrt{2}$. In order to prevent coupling to the
antisymmetric state we need to explicitly break reflection
symmetry along the $x$-direction. This can be achieved by
ramping the chemical potential as illustrated in
Fig.~\ref{Fig:optical_potential}c). If we subsequently increase
the hopping between the two plaquettes ($V_{1x} \rightarrow 0$)
and finally eliminate the shift in the chemical potential
($V_{2x} \rightarrow 0$), we can adiabatically connect the initial state to
the symmetric ground state of two coupled plaquettes. The coupling
transformation is protected by a considerable gap as shown in
Fig.~\ref{Fig:PlaquetteGaps}.
The gaps for the subsequent decoupling process which keeps full
exchange symmetry between the two plaquettes reveal the small
pair binding energy in the vicinity of small interplaquette hopping
$t$ as shown in Fig.~\ref{Fig:PlaquetteGaps} for $U/t=2.5$ (pair binding)
and $U/t=6.0$ (no binding). The low-energy dynamics shown in
Fig.~\ref{Fig:PlaquetteGaps} can be probed by measuring the
structure factor in light-scattering spectroscopy
\cite{LongPaper}.


\begin{figure}[t]
  \includegraphics[width=86mm]{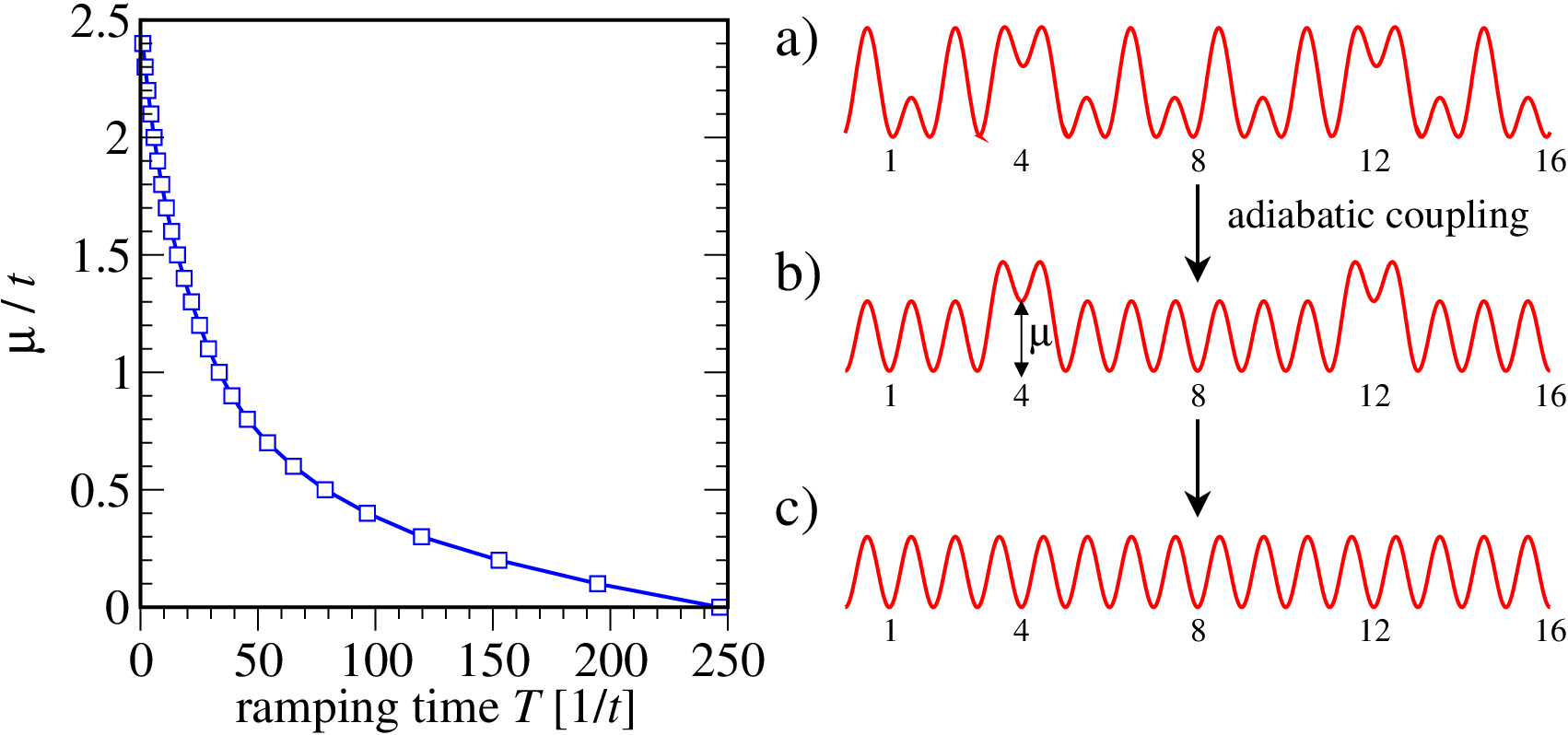}
  \caption{Ramping times needed to achieve a fidelity larger than 0.999 for
                  an adiabatically evolved wavefunction as the chemical potential
                  shift $\mu$ at predefined lattice sites is reduced.
                  The adiabatic protocol is performed in small sequences
                  $\mu \rightarrow \mu - \delta\mu$ with $\delta\mu=0.1$.
                  Doping $\delta=1/8$.
                  On the right: a) decoupled plaquettes; b) comb-like structure;
                  c) ladder geometry. }
  \label{Fig:combramp}
\end{figure}

{\em Doped d-wave RVB ladders --} Finally, we discuss half-filled
ladder systems by coupling multiple plaquettes and the preparation
of a hole doped $d$-wave RVB state. $d$-wave hole pairing in
ladders is strongest for $U/t\approx 2.5$; pairs are localized and
pinned with a period $1/\delta$ for open boundary conditions and
doping $\delta$. To enable a fast adiabatic coupling
process we prepare an initial optical lattice that already places
hole pairs close to their final location in the doped ladder by
mimicking the $1/\delta$ periodicity as shown in
Fig.~\ref{Fig:combramp}b).
Using a pattern loading technique \cite{Rabl:03} we can load a gas
of fermionic atoms into this optical lattice so that the shifted
sites in this comb-like structure remain depleted while all other
rungs in the ladder system are half-filled. When adiabatically
coupling this initial state to the final RVB state of the doped
half-filled ladder by reducing the chemical potential shifts
$\mu$, the time scale is dominated by the time needed to establish
phase coherence between previously unconnected ladder parts.

In our numerical simulations, we considered a ladder with 32 rungs
filled with 56 particles, i.e.\ doping $\delta=1/8$. In the
initial configuration 28 rungs are half-filled and the rungs 4,
12, 20, 28 depleted, see Fig.~\ref{Fig:combramp}b). 
Using adaptive time-dependent density matrix renormalization group 
(DMRG) algorithms \cite{Daley:04}, 
we have calculated the time scales for ramping down $\mu$ at the specified sites.
We find that with an overall ramping time $T
\approx 250/t$, following the protocol shown in Fig.~\ref{Fig:combramp}), the fidelity is larger than 0.9 .
 To improve this bound to $\approx0.99$, we estimate $T \sim 450/t$.



Phase coherence in the ladder RVB state can be experimentally
probed via molecule formation where the molecules are generated
through a laser-induced Raman transition that couples to a
molecular state $m$ with $d_{x^2-y^2}$-symmetry with a transition
matrix element of the form $m^{\dagger}\Delta$ \cite{Buechler:05}.
If the ladder exhibits quasi-long range order the generated
molecules will be phase coherent forming a quasi-condensate.
A truly long-range superfluid ground state of $d$-wave RVB pairs
can be stabilized in experiment by weak coupling of ladders.
Increasing the inter-ladder coupling in the experimental setup
will answer one of the foremost open questions of solid state
physics: is the $d$-wave RVB state on ladders adiabatically
connected to the ground state of the Hubbard model on the uniform
square lattice, and hence the RVB theory of high temperature
superconductivity confirmed -- or is there a quantum phase
transition to a new phase?


ST acknowledges support by the Swiss National Science Foundation.
Research at the University of Innsbruck is supported by the
Austrian Science Foundation and EU projects. We thank the Aspen
Center for Physics and the Kavli Institute for Theoretical Physics
in Santa Barbara, where this work was initiated. Some of the
simulations were based on the ALPS libraries
\cite{ALPS}.


\end{document}